**Title:** Assessing the Habitability of the TRAPPIST-1 System Using a 3D Climate Model.

**Author:** Eric T. Wolf [1]

**Affiliation:**
[1] Laboratory for Atmospheric and Space Physics, Department of Atmospheric and Oceanic Sciences, University of Colorado, Boulder

**Corresponding Author Information:**
Eric T. Wolf
Laboratory for Atmospheric and Space Physics
3665 Discovery Drive
Campus Box 600
University of Colorado
Boulder, CO 80303-7820
eric.wolf@colorado.edu
240-461-8336

*Abstract:*

The TRAPPIST-1 system provides an extraordinary opportunity to study multiple terrestrial extrasolar planets and their atmospheres. Here we use the National Center for Atmospheric Research Community Atmosphere Model version 4 to study the possible climate and habitability of the planets in the TRAPPIST-1 system. We assume ocean covered worlds, with atmospheres composed of $N_2$, $CO_2$, and $H_2O$, and with orbital and geophysical properties defined from observation. Model results indicate that the inner three planets (b, c, and d) presently reside interior to the inner edge of the traditional liquid water habitable zone. Thus if water ever existed on the inner planets, they would have undergone a runaway greenhouse and lost their water to space, leaving them dry today. Conversely, the outer three planets (f, g, and h) fall beyond the maximum $CO_2$ greenhouse outer edge of the habitable zone. Model results indicate that the outer planets cannot be warmed despite as much as 30 bar $CO_2$ atmospheres, instead entering a snowball state. The middle planet (e) represents the best chance for a presently habitable ocean covered world in the TRAPPIST-1 system. Planet e can maintain at least some habitable surface area with 0 – 2 bar $CO_2$, depending on the background $N_2$ content. Near present day Earth surface temperatures can be maintained for an ocean covered planet e with either 1 bar $N_2$ and 0.4 bar $CO_2$, or a 1.3 bar pure $CO_2$ atmosphere.

*1. Introduction*

Recently seven planets were found orbiting the ultracool star TRAPPIST-1 in a transiting configuration (Gillon et al. 2016; Gillon et al. 2017). These seven planets are remarkable because they are all terrestrial sized, with masses ranging from $0.41 - 1.38$ $M_\oplus$ and radii ranging from $0.755 - 1.086$ $R_\oplus$. These planets receive relative incident stellar fluxes of $0.131 - 4.24$ $S/S_\oplus$, where $S_\oplus$ is the total stellar flux received by the modern Earth (~1360 $Wm^{-2}$). The transiting configuration of these planets means that current and future missions can attempt to characterize their atmospheres through transit spectroscopy. Thus the TRAPPIST-1 system will provide the community new ground where theory on the evolution of terrestrial planet atmospheres can be tested against observations. Gillon et al. (2017) suggest that the seven TRAPPIST-1 planets may be "temperate" because all have equilibrium temperatures below ~400 K. However, equilibrium temperature is a rudimentary metric of planetary climate. Equilibrium temperature ignores the greenhouse effect, while the albedo is generally unknown. To obtain an improved assessment of habitability for the TRAPPIST-1 planets, one must use advanced climate models that can adequately compute the greenhouse effect, planetary albedo, and ultimately the surface temperature.

Earth provides the only archetype for a robustly habitable world. Thus by definition, habitable planets must maintain generically Earth-like surface conditions, with abundantly available liquid water (Hart 1979). The necessary condition of surface liquid water implies a surface temperature range of 273 K − 373 K. However, despite our Earth-centric definition for planetary habitability, TRAPPIST-1 and its planetary system are quite different from our own. TRAPPIST-1 is an ultracool dwarf star, meaning that it is small, dim, and red. Each of these factors has important consequences for the potential climates of its seven orbiting planets. TRAPPIST-1 has a mass of only 0.08 $M_\odot$ and a luminosity of only $5.25 \times 10^{-4} L_\odot$ (Gillon et al. 2016). Thus, despite all seven planets orbiting within 0.063 AU from the star, they receive a moderate range of incident stellar radiation. However, their close-in orbits mean that all planets in the system are likely locked into tidal synchronization, particularly given their low eccentricities (Gillon et al. 2017). Thus one side of the planet always faces the star, and the planetary rotational period equals the orbital period. Generally, tidal locking implies planetary rotation rates that are slower than Earth's, and this holds true for the TRAPPIST-1 system. However, the orbital periods for planets b − f are less than ~10 Earth days. Thus even if synchronously rotating, the Coriolis effect will be non-negligible and these worlds retain zonal flow patterns (Kopparapu et al. 2016). TRAPPIST-1 is quite cool, with an effective temperature of only ~2560 K. Thus its emitted stellar radiation is shifted towards the near infrared compared with our Sun. This shift affects radiative interactions in the atmosphere and with the surface, because near infrared radiation is more readily absorbed by water vapor and sea ice (Shields et al. 2013). With these characteristics of the star-planet system in mind, we conduct 3D climate calculations for planets in the TRAPPIST-1 system assuming atmospheres composed of $N_2$, $CO_2$, and $H_2O$, following the traditional assumptions for terrestrial planetary atmospheres within the habitable zone (Kasting et al. 1993; Selsis et al. 2007; Kopparapu et al. 2013).

*2. Methods*

We use a modified version of the Community Atmosphere Model (Neale et al. 2010) version 4 (CAM4) from the National Center for Atmospheric Research. We have used this particular model version previously for studying a variety of Earth-like atmospheres (Wolf & Toon, 2013; 2014a; 2014b; 2015; Wolf et al. 2017), and CAM4 has been frequently used elsewhere for studying slow rotating planets around M-dwarf stars (Yang et al. 2013, 2014; Kopparapu et al. 2016; Wang et al. 2016). We have modified the radiative transfer code in the model (*e.g.* Wolf & Toon, 2013), and have also incorporated new methods to improve the numerical stability of the model for these exotic atmospheres. We use the planetary masses, radii, and surface gravity determined from observations (Gillon et al. 2017, table 1). We assume that all planets are locked into synchronous rotation (*i.e.* a 1:1 spin-orbit ratio), thus their rotational period is equal to their orbital period. We assume a completely ocean covered planet, with zero ocean heat transport within a 50-m thermodynamic slab ocean (Bitz et al. 2012). Sea ice forms wherever the sea surface temperature falls below the freezing point of seawater (-1.8º C in the model). We use 4º × 5º horizontal resolution with 40 vertical levels extending from the surface up to a 1 mb model top. Clouds and convection use the subgrid scale parameterizations of Rasch and Kristjánsson (1998) and Zhang and McFarlane (1995) respectively. We use the incident stellar spectra from the BT_Settl stellar models for a 2600 K star (Allard et al. 2003, 2007). We assume several basic atmospheric compositions; $N_2 + H_2O$, $N_2 + CO_2 + H_2O$, and finally $CO_2 + H_2O$. We conduct simulations for planets d, e, and f, which lie at the center of the TRAPPIST-1 system. As described below, the habitability of planets b and c can be inferred from results for planet d. Likewise, the habitability of planets g and h can be inferred from results for planet f.

*3. Results*

Figure 1 shows time series of 3D climate model results for global mean surface temperature ($T_s$) and top-of-atmosphere energy balance for simulations of planet d. Planet d receives an incident stellar flux of 1.143 $S/S_\oplus$, with a 4.05 Earth-day period. With 1 bar $N_2$ and no greenhouse gases (other than $H_2O$), the climate undergoes a thermal runaway and becomes uninhabitable. In Figure 1, simulations were run for 40 years. At that point $T_s$ ~ 380 K, the maximum temperature of the atmosphere exceeds ~450 K, and a large (~40 $Wm^{-2}$) residual top-of-atmosphere energy imbalance remains, indicating further warming would occur if the simulation continued. Note also, when $T_s$ ~ 380 K, the total atmospheric pressure has doubled, because now the atmosphere contains ~1 bar of $H_2O$ in addition to its dry constituents. These water dominated atmospheres could lose an Earth-ocean of water to space in only several million years at the diffusion-limited rate (Hunten 1973). Thus, planet d is most likely hot, dry, and uninhabitable today.

We do not explicitly simulate planets b and c here. They receive stellar fluxes of 2.27 and 4.25 $S/S_\oplus$ respectively, and thus they would be significantly hotter than planet d given identical atmospheric compositions. Thus planets b, c, and d reside interior to the traditional liquid water habitable zone. This diagnosis is in agreement with the habitable zone limits of Kopparapu et al. (2013) for low mass stars, and also preliminary assessments of the system by Gillon et al. (2017). However, some studies suggest that locally habitable conditions may exist for dry (*i.e.* water-limited) planets that lie interior to the traditional habitable zone (Abe et al. 2011; Leconte et al. 2013a).

Conversely, simulations of planet f cannot be prevented from entering a completely ice covered state despite dense $CO_2$ atmospheres (Figure 2). Planet f receives only 0.382 $S/S_\oplus$ with a 9.21 Earth-day period. Here we find that even with 30 bars of $CO_2$, planet f would be completely ice covered. Furthermore, for all simulations of planet f (Figure 2), temperatures become cold enough that $CO_2$ would condense onto the surface, and thus these atmospheres would collapse. Planets g and h, receive 0.258 and 0.131 $S/S_\oplus$ respectively. While we do not explicitly simulate these worlds, they receive considerable less stellar flux than does planet f, and thus they too would be unable to escape a snowball state if warmed by $CO_2$ alone. Thus we conclude that planets f, g, and h lie outside the traditional liquid water habitable zone defined by the maximum $CO_2$ greenhouse limit.

Note that the 1D modeling study of Kopparapu et al. (2013) suggests that planets f and g may fall within the maximum $CO_2$ greenhouse limit, however they make several assumptions in their model that lead to warmer planets. First they neglect increases to the surface albedo due to expanding sea ice and snow. Second, they ignore increases to the planetary albedo due to the formation of thick substellar clouds on synchronous rotators (Yang et al. 2013; Kopparapu et al 2016). Finally, they assume that the atmosphere is saturated with respect to water vapor, artificially maximizing the greenhouse effect. Thus, the Kopparapu et al. (2013) results may be overly generous with respect to the maximum $CO_2$ greenhouse limit for the outer edge of the habitable zone. However, others have suggested that $H_2$ could play a significant role in warming planets at low stellar fluxes (e.g. Pierrehumbert & Gaidos 2011).

Planet e, the central planet in the system, provides the most viable candidate for a robustly habitable world. Figure 3a shows results for the global mean surface temperature of planet e, for simulations with a 1 bar $N_2$ background and varying $CO_2$ (red), and also for pure $CO_2$ atmospheres (blue). Figure 3b shows the global mean percent sea ice coverage. Figure 3c shows the percent of habitable surface area. The habitable area is defined as the percent of the planet's surface that is both ice free, and has a surface temperature less than 310 K. While some life forms on Earth survive at hotter surface temperatures or in glaciated areas, this range of surface conditions (approximately) encompasses the limits for human biological functioning unaided by technology (Sherwood and Huber 2010).

Simulations indicate that planet e can maintain habitable surface conditions for a variety of atmospheric compositions. Planet e receives only 0.662 $S/S_\oplus$ with a 6.10 Earth-day period. Thus, without additional greenhouse gases planet e would be cold. However, even with thin atmospheres planet e can remain habitable locally at the substellar point. With 1 bar $N_2$ and zero $CO_2$ (not shown in Figure 3), $T_s \sim 227$ K, and a small part of the ocean (~13%) remains thawed immediately around the substellar point, with temperatures hovering near ~280 K locally. Marginally warmer conditions are found with an Earth-like composition (1 bar $N_2$ + $10^{-4}$ bar $CO_2$), and also in the case of a thin pure $CO_2$ atmosphere with a surface pressure of only 0.25 bar. These cases have $T_s \sim$ 240 K, sea ice coverage of ~80%, while ~20% of the planet surface is habitable.

Global mean surface temperatures near those of present day Earth (~288 K) can be maintained on planet e presently with either 1 bar $N_2$ + 0.4 bar $CO_2$, or similarly by 1.3 bar $CO_2$. Perhaps coincidentally, an Earth-like temperature coincides with maxima in the habitable surface area (>95%) with ice confined to the poles and moderate surface temperatures elsewhere. However, for further increasing $CO_2$ amounts, the habitable surface area sharply declines as surface temperatures locally warm beyond the human heat stress limit (Sherwood & Huber 2010). The habitable area eventually falls to zero for atmospheric compositions of 1 bar $N_2$ + 2 bar $CO_2$, and similarly for 4 bar $CO_2$. These hot but stable states have $T_s \geq 330$ K. However, their stratospheric $H_2O$ volume mixing ratios remain small (~$10^{-5}$ at 1 mb) due to efficient cold trapping caused by cooling of the middle and upper atmosphere from high $CO_2$ concentrations (e.g. Wordsworth & Pierrehumbert, 2013). Thus while the runaway cases described in Figure 1 would cause planets b, c, and d to be desiccated today, planets with hot stables climates shown in Figure 3 could retain their water for long periods of time.

Figure 4 shows the temporal mean surface temperature, cloud water column, net outgoing thermal radiation, and reflected stellar radiation from the primary atmospheric states studied here: a completely glaciated "snowball" planet f, a "cold" but marginally habitable planet e, a "temperate" planet e at modern Earth temperatures, a "hot" and uninhabitable planet e, and finally an incipient thermal "runaway" for planet d. Descriptions of each simulation are in the left margin of Figure 4. In surface temperature contour maps, solid white lines indicate the sea ice margin and dashed white lines indicate where $CO_2$ would condense onto the surface of the planet. The substellar point is located at the center of each frame. Note that the runaway case will continue to increase in temperature beyond what is illustrated, and would eventually lose its water to space. Thus the image of a runaway shown is a snapshot of a transient state.

The snowball planet f is completely covered in ice, and has minimal clouds and water vapor in its atmosphere. The thermal emitted flux is low due to its cold temperature, but the reflected stellar energy is significant due to snow and ice cover. Though not explicitly included in the model, the atmosphere is cold enough that $CO_2$ would condense onto the night side of the planet, causing the atmosphere to collapse. The cold planet e can maintain open ocean only immediately around the substellar point, but the majority of the planet is ice covered. Thick clouds form over open waters at the substellar point, and contribute significantly to the planetary energy balance by increasing the albedo and decreasing the emitted thermal flux where clouds are thickest. This pattern is also seen in warmer cases. For the cold planet e, sea ice also contributes to the reflected stellar energy near the terminators. A temperate planet e is the most favorable scenario, and maintains habitable conditions over virtually its entire surface. Sea ice is confined to the poles. Clouds are thick over the substellar point and poles, and the thermal and reflected flux fields mirror the distribution of the clouds. The final two states are increasingly hot and uninhabitable. As climate warms, surface temperatures become uniform across the planet and sea ice vanishes entirely. Despite significant water vapor in their atmospheres, relative humidity and clouds decrease for hot atmospheres. For increasing temperatures, the day-side becomes increasingly dry (*i.e.* low relative humidity) and substellar clouds thin and eventually vanish. In the runaway case, clouds can only be maintained on the night-side and along the terminator. The reduction in day-side clouds reduces the amount of reflected stellar energy from these hot worlds. The outgoing thermal flux for the hot case is comparable to that of the temperate case, however the outgoing thermal flux becomes large for an atmosphere in runaway.

The net outgoing thermal and reflected stellar flux maps shown in Figure 4 begin to tell us how these climate states may appear to the distant observer. From these flux maps, it is helpful to construct phase curves (*e.g.* Koll & Abbot 2015, appendix C) of the thermal emitted flux and the planetary albedo (Figure 5). Note that at a phase angle of 0º the observer sees the day-side of the planet, and at phase angles of -180º and 180º the observer sees the night-side of the planet. From phase curves, one can distinguish between atmospheres of interest. An incipient thermal runaway emits between 300 − 400 Wm$^{-2}$ of thermal flux, due to its hot and sub-saturated atmosphere. However, its albedo is small and near zero at a phase angle of 0º. The thermal emitted flux and albedo are out of phase, with maximum emission occurring at a phase angle of 0º concurrent with the minimum in albedo. The albedo increases near -180º and 180º due to grazing incidence scattering from clouds found along the terminator regions. Note that grazing incidence causes the albedo to increase at phase angles near -180º and 180º for all cases.

The snowball case can be distinguished by having a low emitted thermal flux (<100 Wm$^{-2}$), while also featuring the highest albedo. We find the albedo to range between 0.3 − 0.5. Temperate, and hot climates are more difficult to distinguish from each other. Their emitted thermal phase curves are virtually identical (~200 Wm$^{-2}$) despite a ~40 K difference in $T_s$. The albedo of the hot climate is about ~20% lower than that of the temperate climate. Finally the cold case emits slightly less thermal flux (~150 – 200 Wm$^{-2}$), with a maximum found at a phase angle of 0º and is out of phase with the hot and temperate cases. The albedo of the cold case is significantly greater that of both hot and temperate cases, however it is less than the snowball case.

*4. Conclusions*

Here we have used a state-of-the-art 3D climate system model to take a first cut at evaluating the climate and habitability of the TRAPPIST-1 system. Planets b, c, and d are likely too hot to support abundant liquid water at their surfaces. These planets would have undergone a runaway greenhouse process, and have probably lost their water to space long ago. Planets f, g, and h are likely too cold to support surface liquid water. If these planets contain water they are probably encased in ice and snow, despite as much as 30 bars of $CO_2$. Planet e is the best chance for a presently habitable ocean covered world in the TRAPPIST-1 system. Planet e can maintain at least some habitable surface area under a variety of atmospheric compositions. With a 1 bar $N_2$ background, planet e is habitable for $CO_2$ amounts up to 1 bar. For pure $CO_2$ atmospheres, planet e is habitable with $CO_2$ ranging from $0.25 - 2$ bars. Planet e can maintain near present day Earth surface temperatures with a 1.3 bar pure $CO_2$ atmosphere, or 1 bar $N_2$ + 0.4 bar $CO_2$.

However, these simulations are predicated on the assumption that each planet has abundant water at their present time and location in the system. The super-luminous pre-main sequence phase of low mass stars may spell doom for planets orbiting in their habitable zones today (Luger & Barnes 2015). Ultracool dwarf stars may take up to ~1 Gyr to settle onto the main sequence, subjecting any planets with intense stellar radiation, driving them into runaway greenhouse conditions. Bolmont et al. (2016) predict that planet d, given its confirmed location in the system at 0.021 AU, may have lost up to ~7

Earth oceans of water. While planet e had not been identified at the time of the study, the Bolmont et al. (2016) results indicate that planet e may have lost several Earth oceans of water during the pre-main sequence phase. Thus planet e would have needed an initial water inventory at least several times greater than the Earth presently for it to retain abundant water today. An alternative idea is that the TRAPPIST-1 planets may have formed further out and then migrated to their current positions (Terquem & Papaloizou 2007), thus circumventing the pre-main sequence runaway phase, and making it easier from them to retain primordial volatiles. Additionally, it has been suggested that surface-mantle volatile cycling may not reach equilibrium until several Gyr after planetary formation, and thus surface water could be replenished from the interior after the pre-main sequence phase concludes (Komacek & Abbot, 2016).

It is also important to note that these simulations originate from a single 3D climate system model. Differences amongst climate models exist, particularly for exoplanetary problems that push the boundaries of these originally Earth-centric codes (*e.g.* Leconte et al. 2013b; Wolf & Toon 2015, Popp et al. 2016). Gillon et al. (2017) assert that 3D climate model simulations of terrestrial planets around low mass stars (*e.g.* Turbet et al. 2016) indicate that planets b, c, and d would undergo a runaway greenhouse, while planets e, f, and g could be habitable given "Earth-like" atmospheres. While our results agree regarding planets b, c, and d, in this study planets f and g are too cold to be habitable. Surely, future works will examine the climate and habitability of the TRAPPIST-1 planets using a variety of 1D and 3D atmosphere models. Through careful model intercomparison, we can gain confidence in our ability to simulate the climates of the TRAPPIST-1 system.

E.T.W. acknowledge support from NASA Planetary Atmospheres Program award NNX14AH17G, and from the NASA Habitable Worlds program award NNX16AB61G. This work utilized the Janus supercomputer which is supported by the National Science Foundation (award CNS-0821794) and the University of Colorado at Boulder. This work was also facilitated through the use of advanced computational, storage, and networking infrastructure provided by the Hyak supercomputer system, supported in part by the University of Washington eScience Institute. E.T.W. thanks the NASA Astrobiology Institute's Virtual Planetary Laboratory at the University of Washington for granting computing time on Hyak. E.T.W. also thanks D. S. Abbot, R.K. Kopparapu, J. Haqq-Misra, and O.B. Toon for helpful comments and discussions.

*7. Figure Captions*

**Figure 1**: Time series model outputs from simulations of TRAPPIST-1d with atmospheric compositions of 1 bar $N_2$ and 1 bar $N_2$ + 0.01 bar $CO_2$. The top panel (a) shows the mean surface and maximum atmosphere temperatures. The bottom panel (b) shows the top-of-atmosphere energy imbalance.

**Figure 2:** Time series model outputs from simulations of TRAPPIST-1f with dense $CO_2$ atmospheres. The top panel (a) shows the mean surface temperatures. The bottom panel (b) shows percent sea-ice coverage.

**Figure 3:** Simulations of TRAPPIST-1e with a various atmospheric compositions. All simulations shown are in equilibrium. Shown are the mean surface temperature (a), the sea ice coverage (b), and the habitable surface area (c). Red lines indicate simulations containing a 1 bar $N_2$ background, plus additional $CO_2$. Blue lines indicate simulations containing a pure $CO_2$ atmosphere. All simulations include $H_2O$.

**Figure 4:** Contour plots of surface temperature, cloud water column, net outgoing thermal flux, and reflected stellar flux for several atmosphere types, including snowball, cold, temperate, hot, and runaway. Note the description of each simulation in the left-hand margin of the figure. In the surface temperature maps, a white solid line indicates the sea ice margin and a dashed white line indicates $CO_2$ condensation onto the surface.

**Figure 5:** Phase curves of emitted thermal flux and planetary albedo based on the simulations shown in Figure 4.

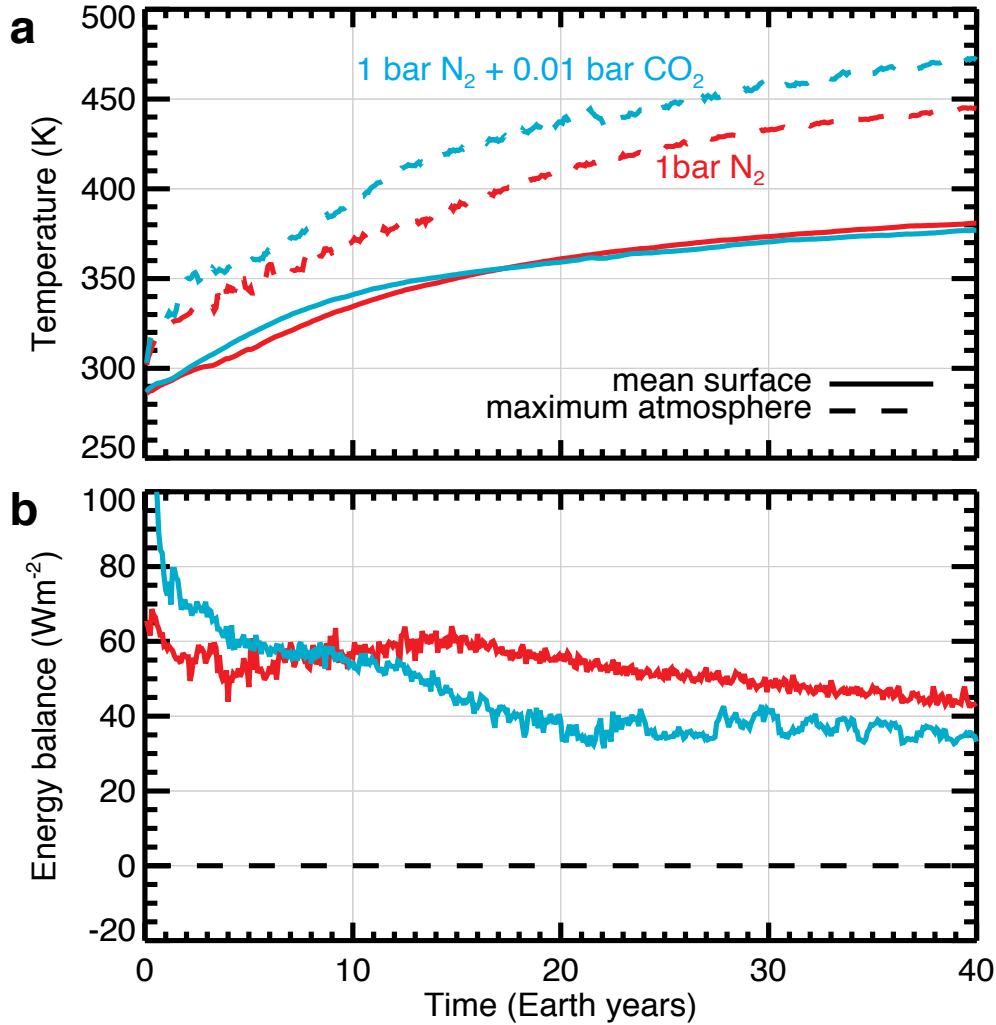

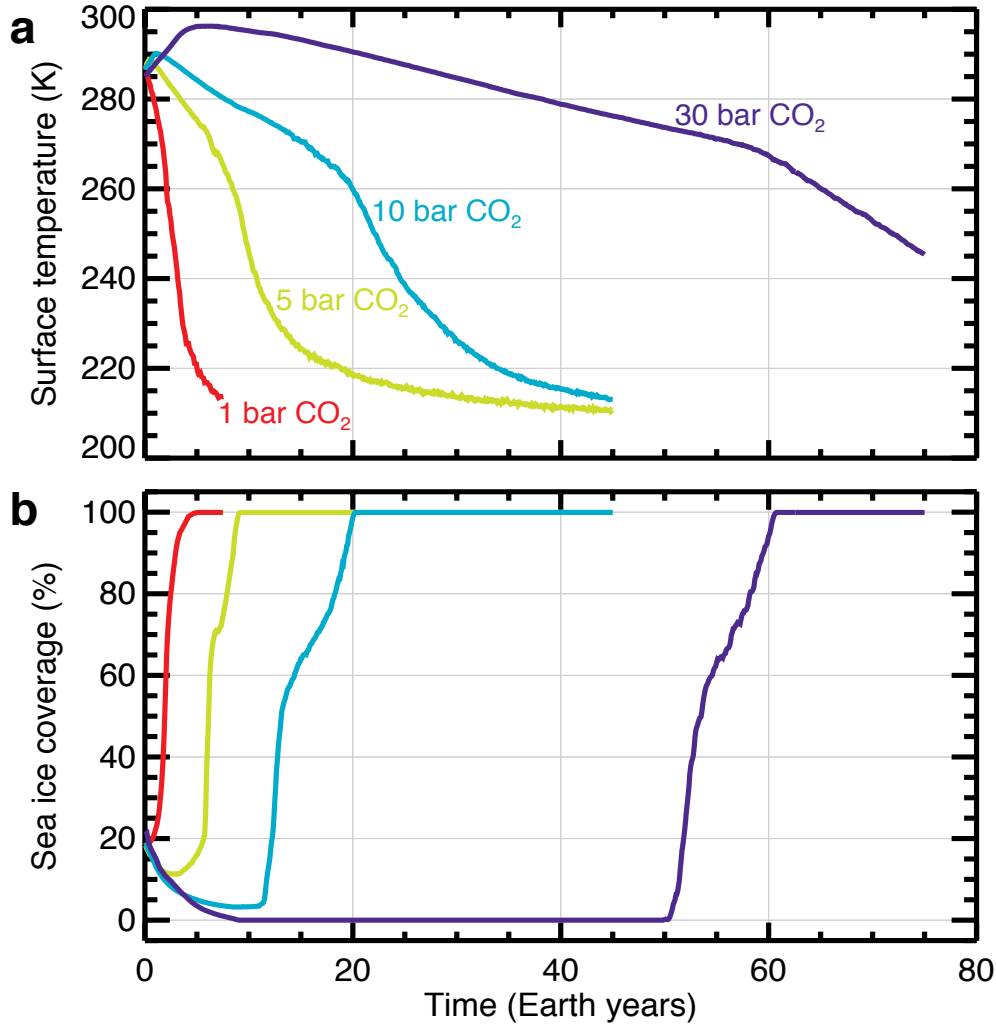

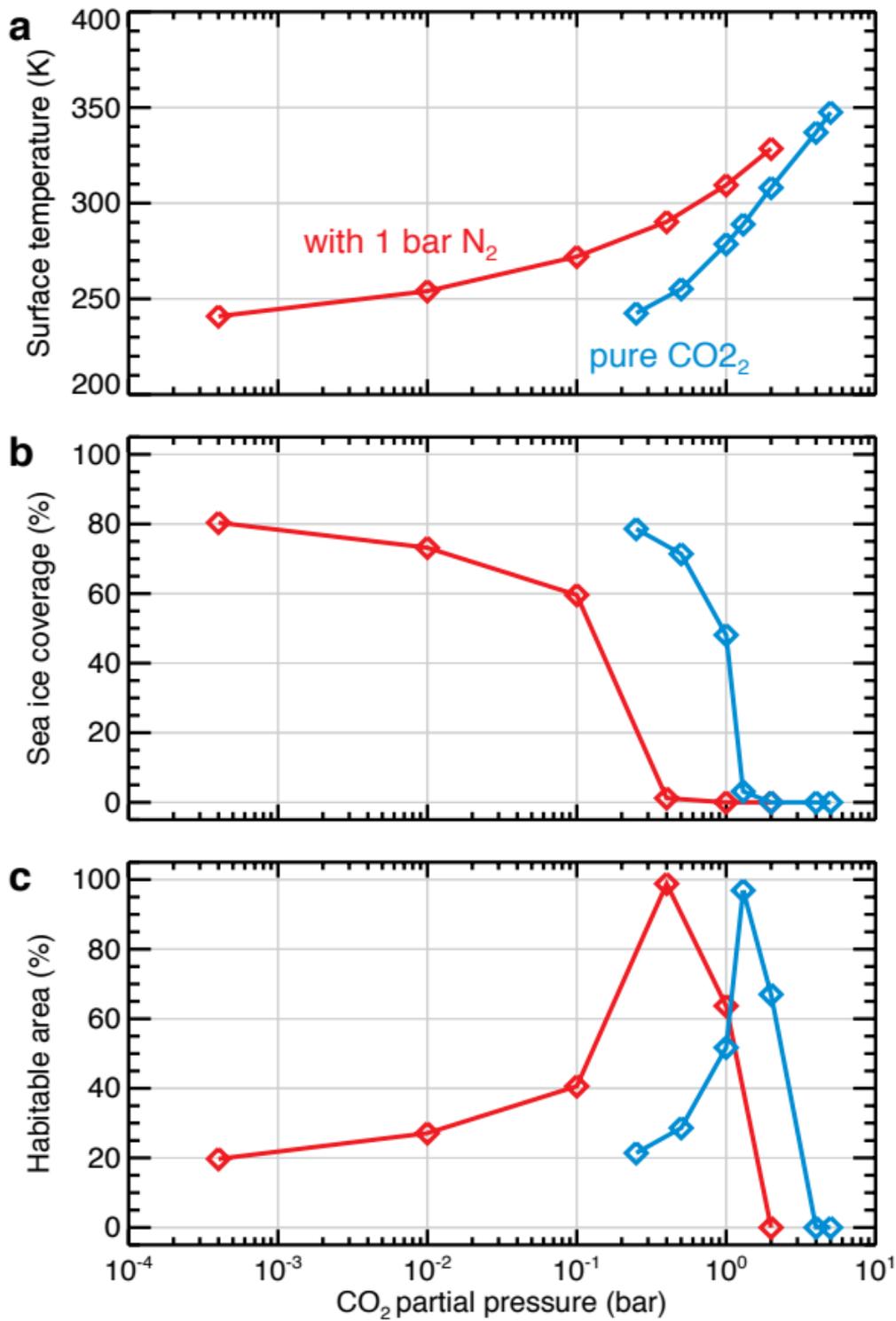

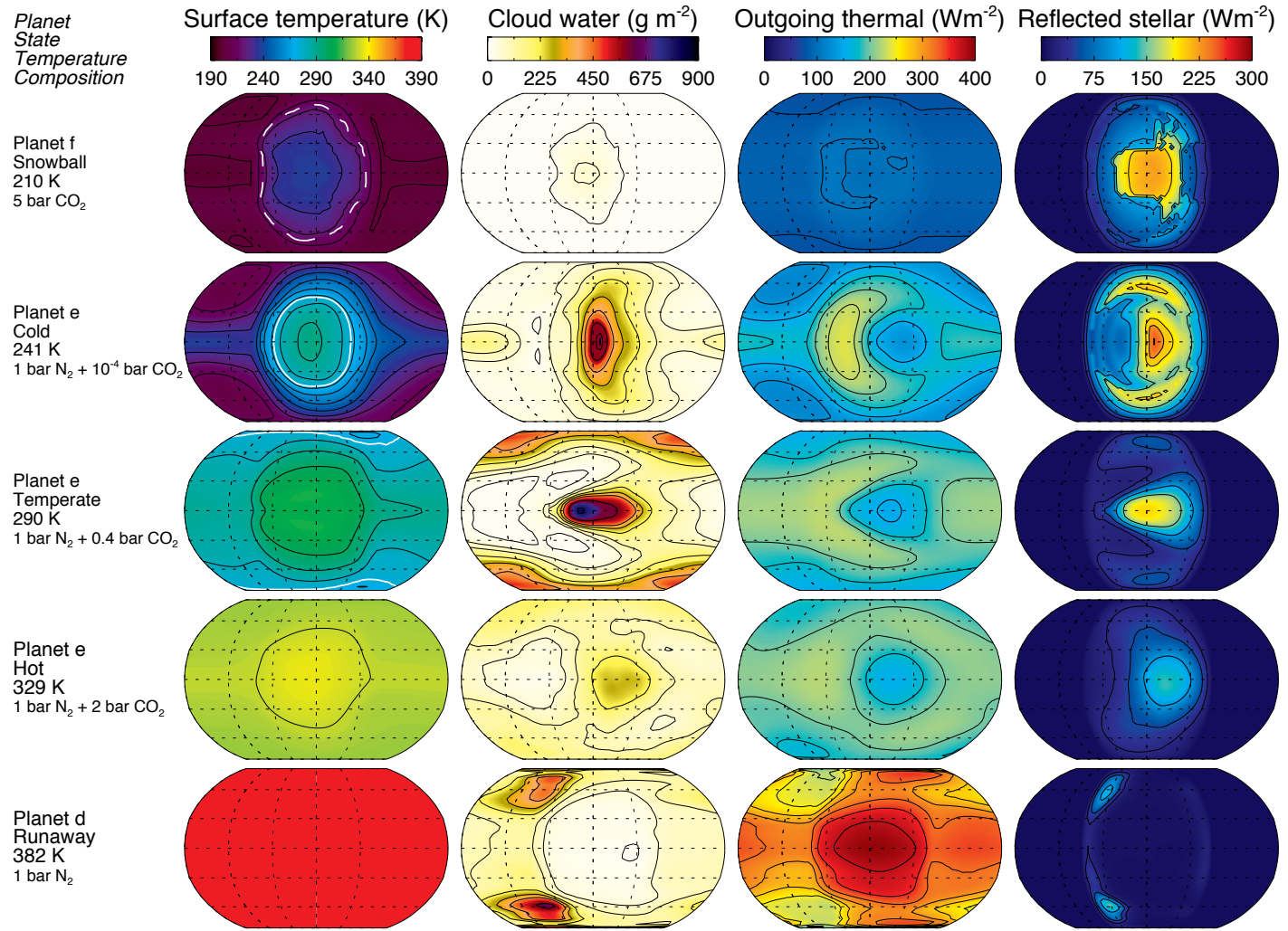

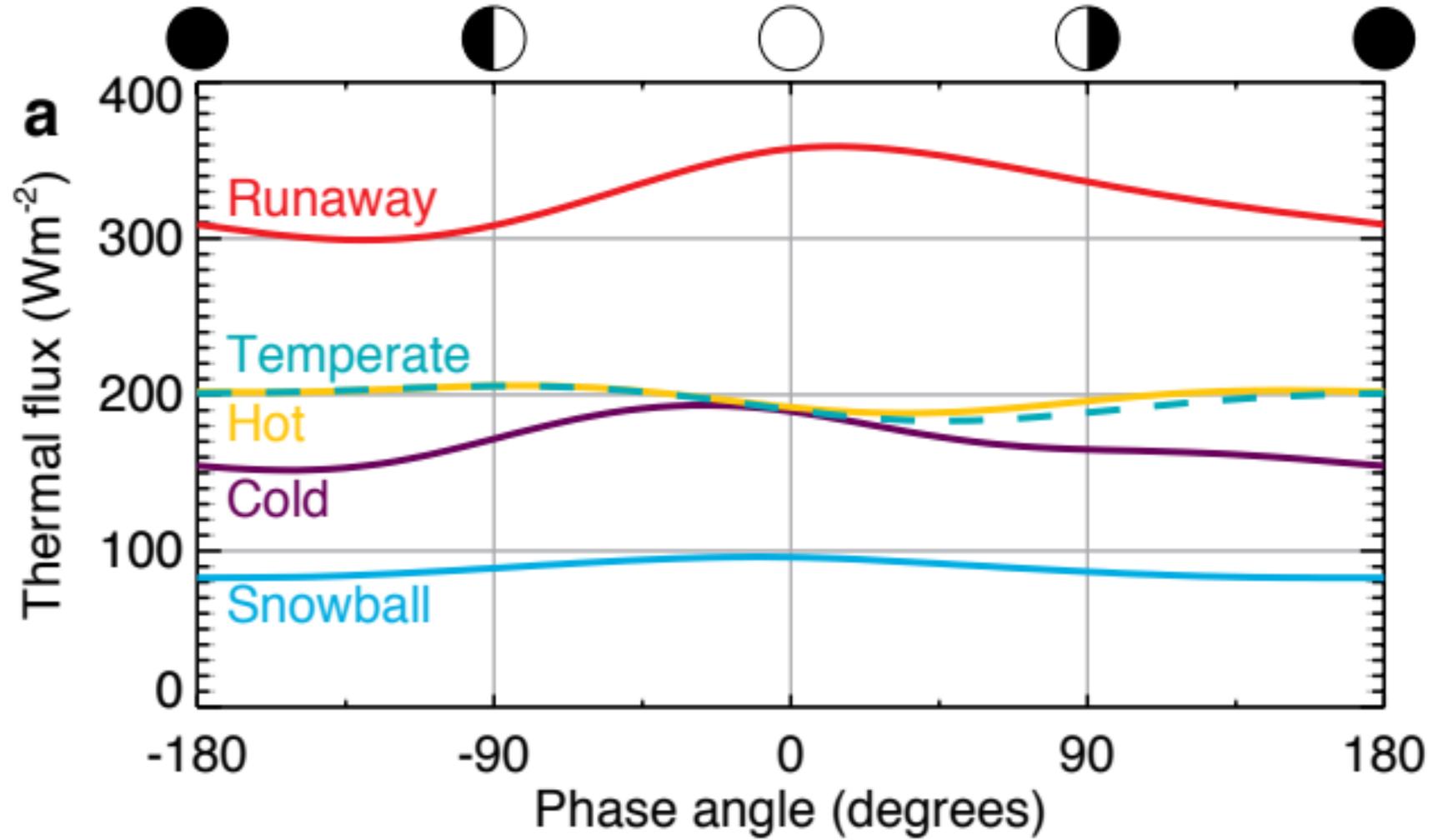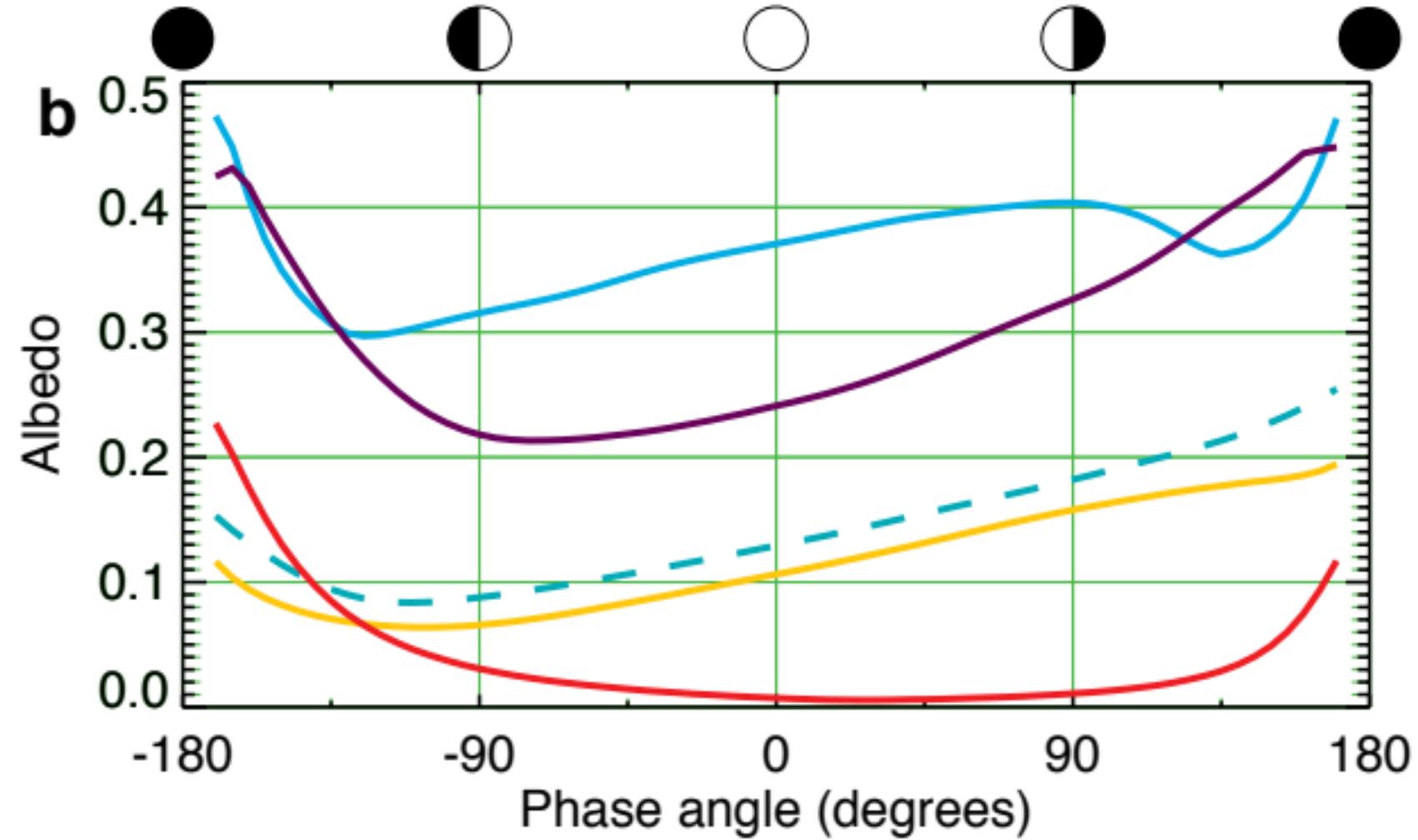

**Title:** Erratum: "Assessing the Habitability of the TRAPPIST-1 System Using a 3D Climate Model." (839:L1 (6pp), 2017)

**Author:** Eric T. Wolf

**Affiliation:**
[1]Laboratory for Atmospheric and Space Physics, Department of Atmospheric and Oceanic Sciences, University of Colorado, Boulder

**Corresponding Author Information:**
Eric T. Wolf
Laboratory for Atmospheric and Space Physics
3665 Discovery Drive
Campus Box 600
University of Colorado
Boulder, CO 80303-7820
eric.wolf@colorado.

Due to a coding error the $CO_2$ continuum absorption was incorrectly applied, affecting simulations with 0.2 bar $CO_2$ and greater. Prior conclusions regarding the habitability of TRAPPIST-1f are now changed. While TRAPPIST-1f with 1 bar $CO_2$ still enters a fully glaciated state, the planet can maintain a global mean surface temperature ($T_S$) of 284 K with 2 bar $CO_2$. A corrected version of Figure 2 is included here. With 5 bar $CO_2$, TRAPPIST-1f maintains $T_s$ = 334 K, in close agreement with results from the LMD Generic Global Climate Model (*personal communication*, M. Turbet). Conclusions regarding TRAPPIST-1e are qualitatively unchanged, however previously reported $T_s$ for simulations with >0.2 bar $CO_2$ were underestimated. A corrected version of Figure 3a is included here. Conclusions regarding TRAPPIST-1d are unaffected. Conclusions regarding thermal and reflected light phases curves yielded from different climate states are also unchanged. Although, the amount of $CO_2$ required to drive different climate states is changed as described in the corrected Figures 2 and 3 included in this erratum. Finally, note that this error is confirmed to be isolated to this work, and was not present in previous works by the author.

**Figure 2:** Time series model outputs from simulations of TRAPPIST-1f. The top panel (a) shows the mean surface temperatures. The bottom panel (b) shows percent sea-ice coverage. With 1 bar $CO_2$ planet f enters a fully glaciated state, however temperatures close to the present day Earth are maintained with only 2 bar $CO_2$.

**Figure 3**: Simulations of TRAPPIST-1e. All simulations shown are in equilibrium. Shown are the mean surface temperatures versus $CO_2$ partial pressure. Red lines indicate simulations containing a 1 bar $N_2$ background, plus additional $CO_2$. Blue lines indicate simulations containing a pure $CO_2$ atmosphere. All simulations also include $H_2O$. Solid lines and diamonds show corrected simulations, while dashed lines indicate previously reported incorrect values.

**Acknowledgements:** E.T. Wolf would like to thank M. Turbet for helpful discussions that aided in tracking down this error

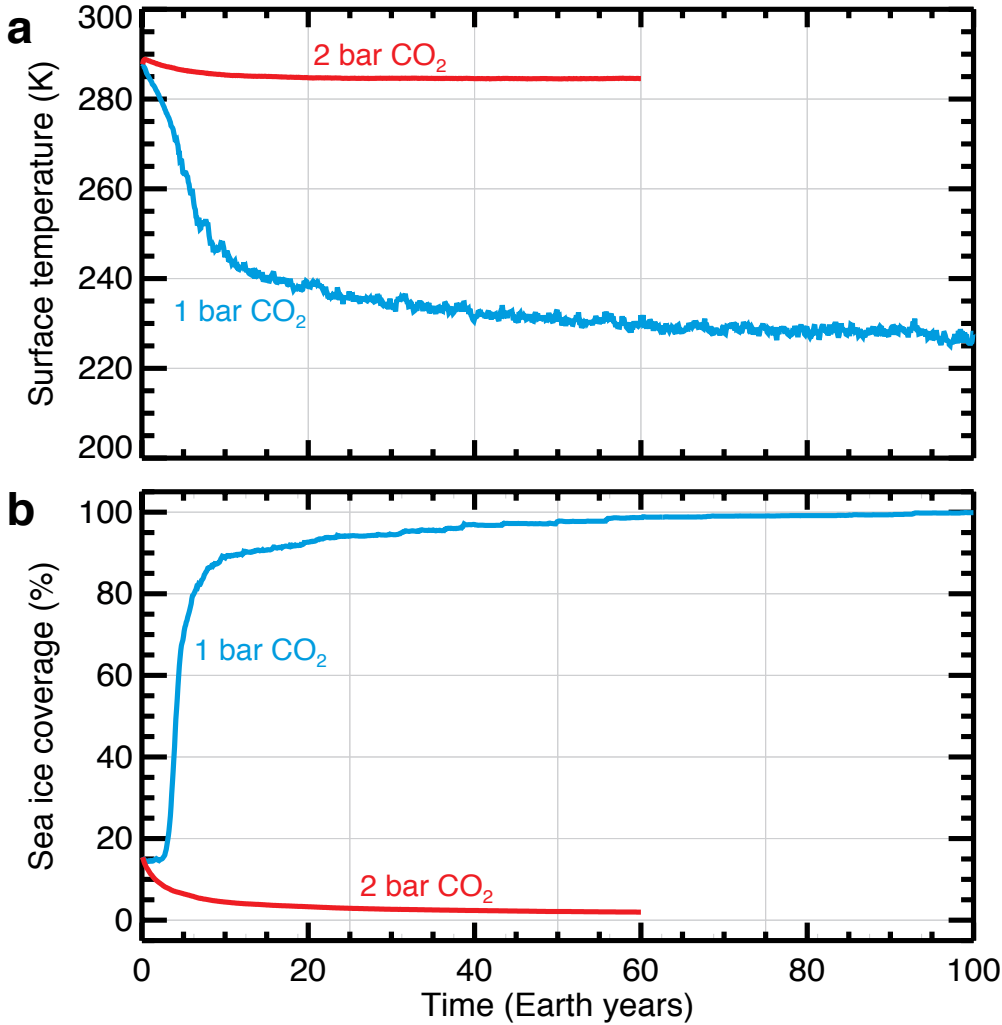

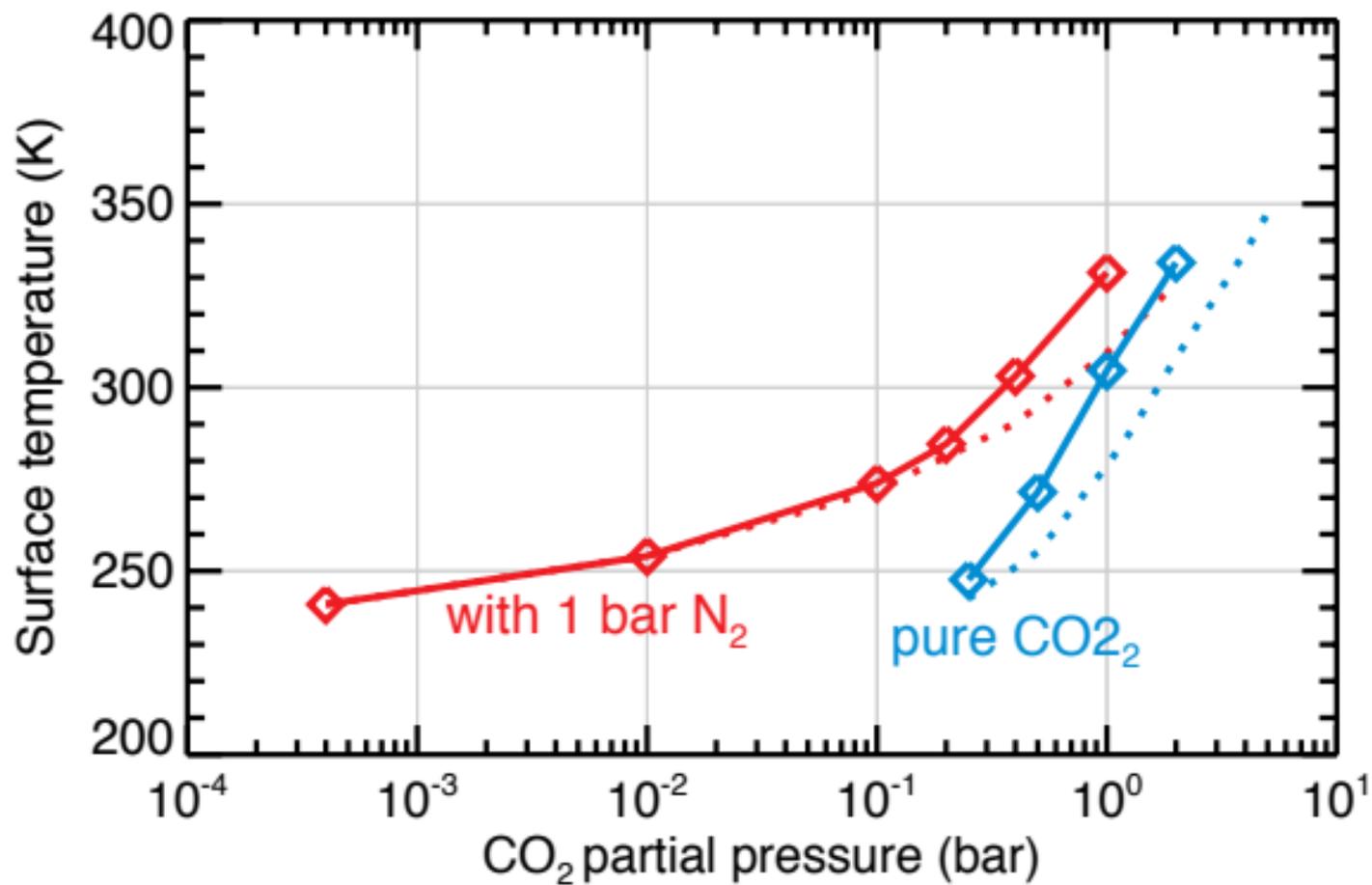

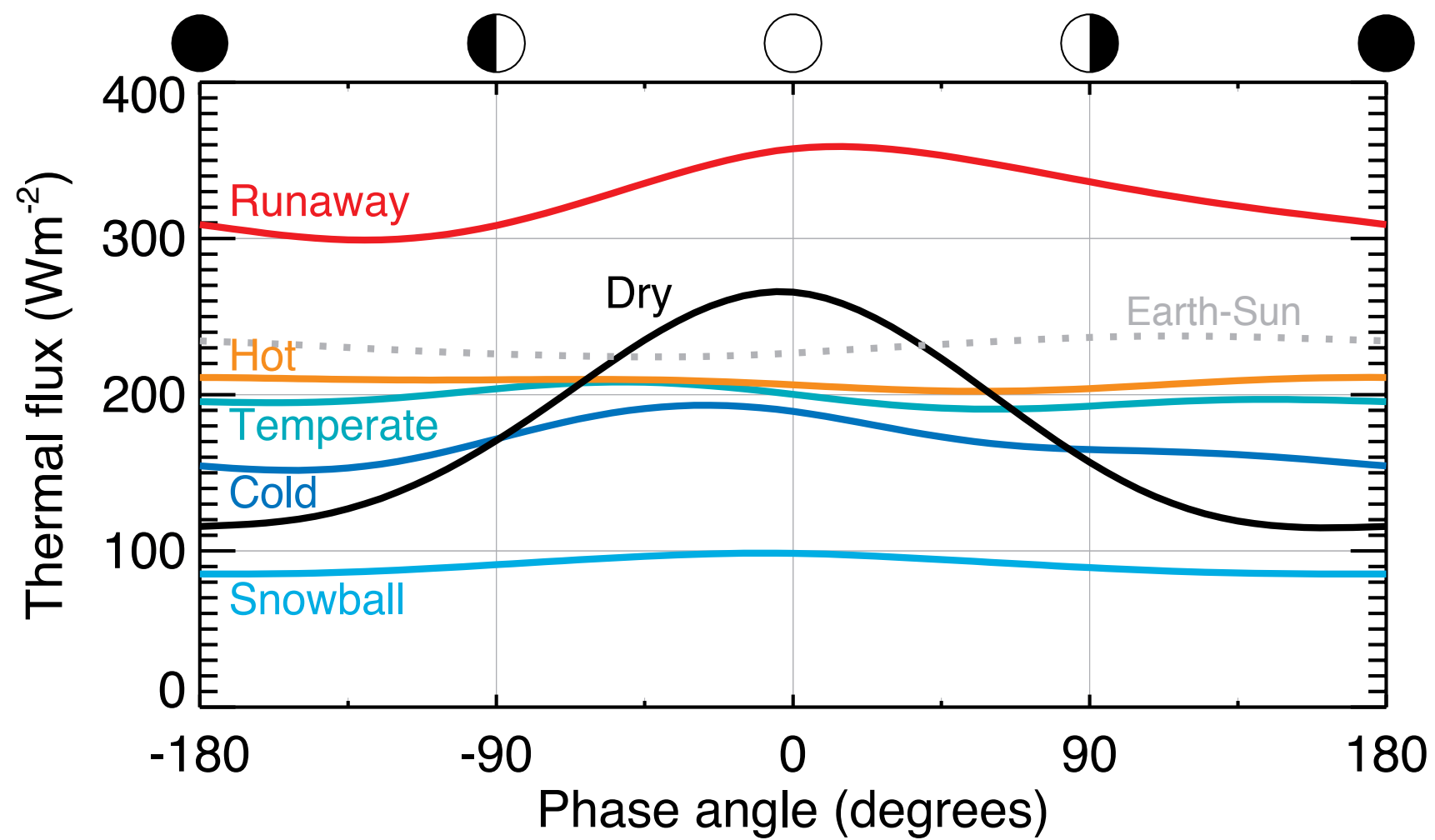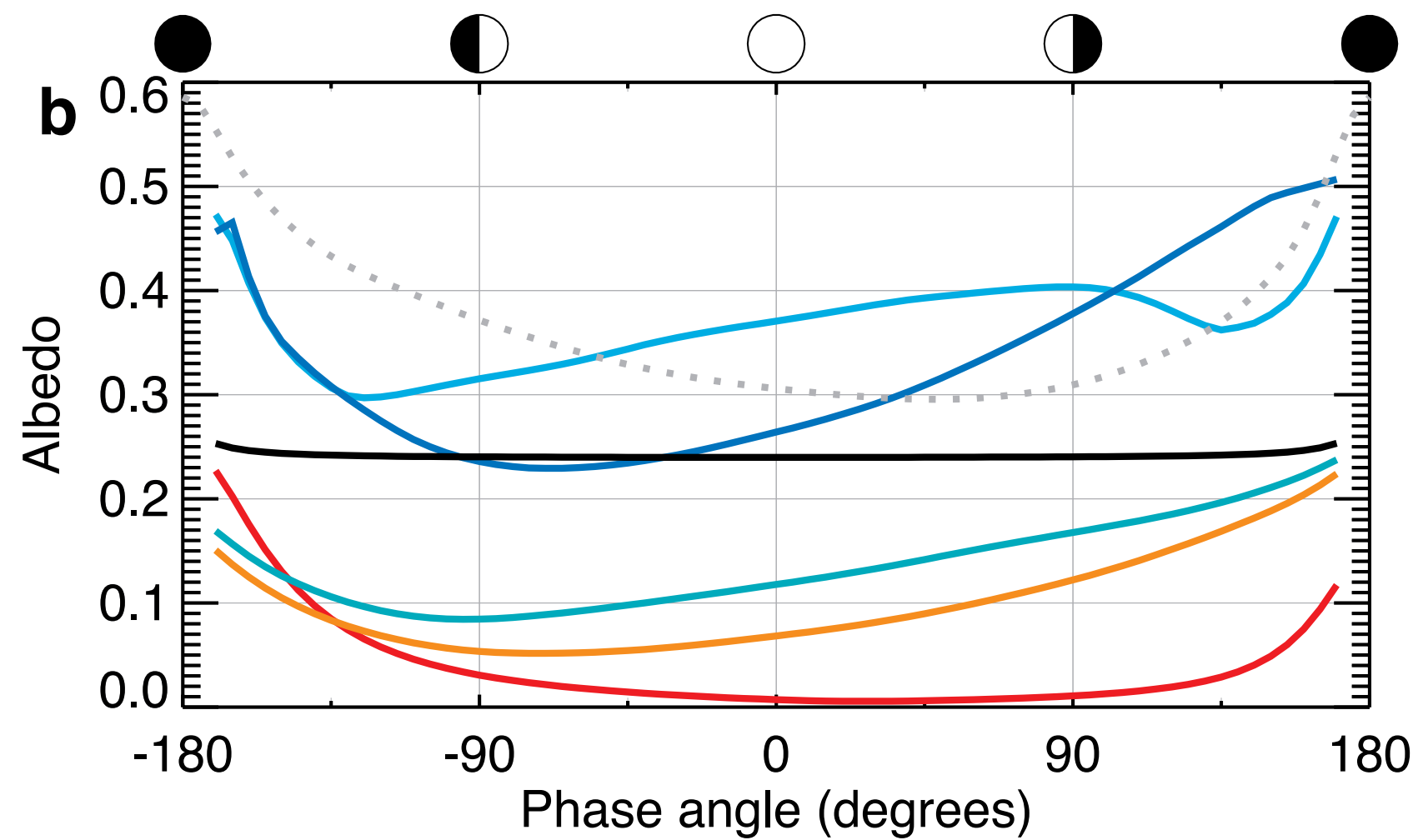